\documentclass[apj]{emulateapj}
\usepackage{natbib}
\begin{document}
\bibliographystyle{apj}

\title{Basic Bell-MHD Turbulence}
\author{Andrey Beresnyak}
\author{Hui Li}
\affil{Los Alamos National Laboratory, Los Alamos, NM 87545}

\begin{abstract}
Nonresonant current instability was identified by Bell (2004) as an important 
mechanism for magnetic field amplification in supernova remnants.
In this paper we focus on studying the nonlinear stage of this instability using 
the incompressible MHD formulation.
We demonstrate that the evolution of magnetic turbulence driven by the Bell instability 
resembles turbulence driven on large scales. More importantly, 
we demonstrate that the energy-containing scale for magnetic fields 
is proportional to the square root of the magnetic energy density. 
Given the observational constraints of the possible field amplification, 
this new relation allows us to directly estimate the maximum
energy of particles scattered by such fields and this estimate is normally below the 
average particle energy. This implies that, without taking into account the feedback
to cosmic rays, the typical scales of Bell fields, in either
linear or nonlinear regime, will be too small to affect high energy particle 
acceleration. We mention several scenarios of back-reaction to cosmic rays that could 
be important. 
\end{abstract}

\maketitle

\section{Introduction}
\def\emf{\mathcal{E}} 

It has long been assumed on energetic arguments that SNRs are responsible for the acceleration of cosmic rays,
at least up to the ``Knee`` of the spectrum at $10^{15}$ eV. 
The most common paradigm proposed for production of cosmic rays in supernova remnants is diffusive shock acceleration (DSA), in which particles
scatter off magnetic turbulence and move across the shock multiple times \citep{krymskii1977,bell1978,blandford1978}.
DSA produces an energy spectrum of $E^{-2}$, which is consistent with $E^{-2.7}$ power law observed at Earth,
accounting for losses due to escape from the Galactic disk.
The scattering rates in an ambient ISM magnetic field, however, are grossly insufficient to provide sizable acceleration and it has been
argued that magnetic fields has to be amplified \citep{voelk1984,blandford1987,malkov2001}. Thanks to the launch of Chandra X-ray Observatory, which had
both the spatial resolution and sensitivity to map X-ray synchrotron emission from TeV electrons, magnetic field amplification
has received observational support. Rather than a diffuse morphology, expected if the magnetic field were low and the
Larmor radius of high-energy electrons were a significant fraction of the SNR diameter, the X-ray synchrotron emission in young SNRs is confined
to a thin region near the shock front. This width, such as in the Cassiopeia A SNR, is interpreted as the post-shock distance traveled by a
$\sim$ TeV electron over a synchrotron cooling timescale \citep{vink2003}, which gives the inferred field strengths of $\sim 100 \mu$G.
Similar analysis in Tycho's SNR gives around 300$\mu$G \citep{cassam2007}. Evidence for magnetic field amplification, however, requires more
detailed physical understanding. 

One of the difficulties in creating a fully self-consistent theory of magnetic field amplification and cosmic ray acceleration
is that it requires treatment of collisionless particles, e.g., CRs, as well as the background magnetized plasma, usually considered as MHD fluid. 
One of the popular approaches to capture the feedback of CRs on the MHD fluid is the so-called streaming instability, where
particles moving faster than Alfvenic speed amplify waves and confine themselves via scattering by the same resonant waves,
see, e.g., \citet{kulsrud1969,lagage1983}. Analytic linear models, however, can not deal with significant magnetic field amplification.
While nonlinear models of streaming instability have already been suggested, e.g., by \citet{diamond2007}, their applicability
is yet to be tested by CR-MHD simulations. In this situation a simplified approach, assuming that cosmic rays are acting on MHD plasma
as an external current and that the bending of cosmic rays is small, was suggested by \citet{bell2004} and quickly gained
popularity \citep[see, e.g.,][]{zirakashvili2008,riquelme2009,bykov2011,rog2012}. Considering cosmic rays as a constant external
current significantly simplifies the problem by bringing it to the realm of pure MHD, where large-scale nonlinear simulations
have been a staple for many years. 

While it is clear that most of the cosmic rays, except the highest energy ones, will be scattered many times over
the lifetime of the supernova remnant, the local short-timescale treatment of cosmic rays as external current may
indeed be useful. Over the years several important properties of Bell's instability have been established on a qualitative
level, e.g., it has been confirmed that the instability continues to grow after going into the nonlinear stage and that the mean scale
of magnetic field also grows. One of the possibilities that has been investigated recently is the growth of large-scale fields
either due to kinetic mechanisms \citep{bykov2011}, which appealed to the oblique instability that would grow slower or on much larger scales
or due to the large-scale dynamo \citep{rog2012}, which would again result in slower exponential growth of the magnetic field on larger scales. 

In this paper we present the study of linear and nonlinear stages of Bell's instability in one of the simplest possible setups
-- incompressible MHD equations driven by constant external current density. The simplicity of our approach highlights
the symmetries of MHD equations and allowed us to achieve a simple physical understanding of the nonlinear stage.
In particular we predict a simple law of growth for the magnetic outer scale, which is of primary importance to determine
if the Bell instability can generate magnetic fields which could scatter high-energy particles. Our answer to the above
riddle seems to be negative. Despite the fact that the magnetic outer scale grows very quickly, so does the magnetic energy density.
Furthermore, they seem to be connected by a universal relation, and, at the levels of magnetic saturation brought
by energetic or observational grounds \citep{bell2004,voelk2005}, we don't expect this magnetic field to be important for scattering.

This paper is organized as follows: Section 1 describes MHD equations driven by external currents and
overviews the global conservation properties and the physical source of energy that drives the turbulence.
Section 2 suggests a simple model for the nonlinear stage of the instability, inspired by extending linear
model to the case when magnetic field is randomly oriented with respect to the external current.
Section 3 presents our numerical results, compares them with the predictions of the model and
determines dimensionless coefficients introduced by the model. Section 4 is devoted to the discussion
of astrophysical implications of the current work.

\section{Bell-MHD equations and conservation laws}
Assuming that cosmic rays have very high energy and do not efficiently interact with the MHD fluid, Bell (2004) suggested that the portion
of the Lorentz force, associated with the cosmic ray current should be subtracted from the total Lorentz force, as this portion is not applied
directly to the fluid. The induction equation, however, is unchanged, as it is the consequence of the Ohm's law. The resulting equations
are MHD equations with an external current, which we call Bell-MHD equations\footnote{Here we used
Heaviside-Alfv\'en units with ${\bf j=\nabla \times B}$, which
avoid having a factors of $\sqrt{4\pi}$, $\rho$ and $c$ in the MHD equations and also expresses the magnetic field in velocity units,
as the density assumed to be unity. It is easy to go back to CGS units remembering that energy density
in Heaviside-Alfven units is $\rho(B^2+v^2)/2$. Also, the current density $j=(j_{\rm CGS}/c)(4\pi/\rho)^{1/2}$.
These units allow us to concentrate on the physics of the instability and keep the narrative compact. They are also
a great choice for units in the numerical studies of incompressible MHD.}:
 
\begin{equation}
(\partial_t+{\bf v}\cdot\nabla){\bf v}=-\nabla P+{\bf j\times B}- {\bf j_e\times B}, \label{dv}
\end{equation}

\begin{equation}
\partial_t{\bf B}=\nabla\times({\bf v \times B}).\label{ind}
\end{equation}

The equations above are basically MHD equations with an extra ``force'' $- {\bf j_e\times B}$.
Physically this extra term means that the MHD fluid has external current embedded in it, which has an electric connection with a fluid,
but does not apply any force to the fluid.
A certain insight into the dynamics could be obtained by reviewing conservation laws for the above system.
Originally, MHD equations have five basic conservation laws for scalar or pseudoscalar quantities: mass $m$, momentum $p$, energy $E$, cross-helicity $H_c$ and magnetic helicity $H_M$.
As the continuity equation is unchanged, it is easy to verify that conservation of mass still holds. Furthermore, as induction equation is unchanged, the magnetic helicity conservation still holds
and multiplying Eq.~\ref{dv} by ${\bf B}$ we can verify that cross-helicity is also conserved. The two conservation laws that are broken by the Bell-MHD system are the energy and 
momentum conservation. The average extra momentum per unit time is related to the average magnetic field ${\bf B_0}$ as $-{\bf j_e\times B_0}$. This extra momentum has
to be provided by external currents. In the case of supernova remnants, the amount of momentum carried by high energy cosmic rays is small compared to the inflowing fluid momentum, therefore
the component of external currents which is perpendicular to ${\bf B_0}$ will be suppressed in one gyration for the current-carrying cosmic rays.
Keeping in mind of this, we will consider only the case when ${\bf j_e \| B_0}$ and the global momentum conservation holds true.
Multiplying Eq.~\ref{dv} by ${\bf v}$ we see that the extra energy per unit time is $-{\bf v\cdot (j_e\times B)}={\bf j_e\cdot (v\times B)}$, i.e. it is associated with the electromotive force (EMF)
of the fluid, ${\bf \emf=v\times B}$, applied to the external current. As we'll see below for the unstable modes, MHD fluid applies such an EMF as to extract energy from the external current.
Obviously, this results in energy
loss in the loop of the external current. In the case of the external current provided by cosmic rays, they are being slowed down by MHD fluid's EMF.
We can also write down Eq. \ref{dv} in Fourier space and investigate energy injection as a function of scale. In order to do this, the equations
for the time derivative of the 
Fourier-transformed velocity $\hat v_k$ have to be multiplied to $\hat v_k^*$, a complex conjugate. The result is the energy injection
with the rate of ${\bf j_e\cdot (v_k\times B_k^*)}$, or
${\bf j_e\cdot \emf_k}$, where ${\bf \emf_k}$ is the power spectrum of EMF, ${\bf (v_k\times B_k^*)}$.

\begin{figure}
\begin{center}
\includegraphics[width=0.8\columnwidth]{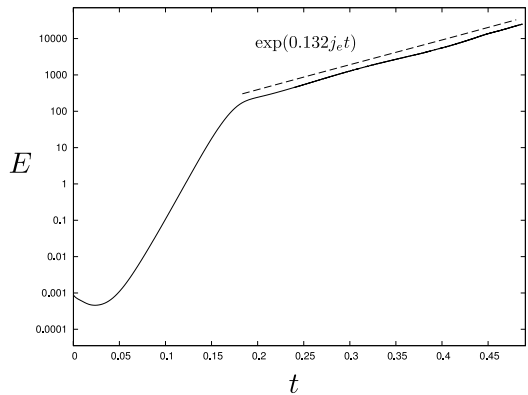}
\end{center}
\caption{Energy grows exponentially in both linear and nonlinear regime of Bell's instability. While the linear growth rate is approximately $j_e$, the nonlinear growth rate is reduced by a factor of $0.132$.}
\label{energy}
\end{figure}

\section{Linear and nonlinear stages}
The linear stage of Bell's instability can be investigated by applying small perturbations to the initial state ${\bf B=B_0}$. This initial state corresponds to plasma current completely canceling out
the external current, i.e. the total current of zero. The mutual repulsion of external and plasma current results in an unstable situation which will grow exponentially from small perturbations
of ${\bf B_1}$ and ${\bf v_1}$. Using linear analysis of the equations above one can verify that the fastest growing mode has a wavenumber
$k_d=j_e/2B_0$  parallel to ${\bf B_0}$, while the 
perturbations ${\bf B_1}$ and ${\bf v_1}$ are perpendicular to ${\bf B_0}$ and have a certain sign of circular polarization, corresponding to the sign
of current helicity ${\bf j_e\cdot B_0}$ \citep{bell2004}.
Energy is equipartitioned between v and B. The fastest growing mode of field perturbations grows
at the rate\footnote{In CGS units $k_d(CGS)=4\pi j_e(CGS)/(2B_0c)$ and the growth rate is
 $(j_e(CGS)/2c)(4\pi/\rho)^{1/2}$.} of $j_e/2$,
while the energy grows at a rate $j_e$.  It is also worth noting that in the presence of scalar viscosity $\nu$ and magnetic diffusivity $\nu_m$ this growth
rate will be decreased by $-(\nu+\nu_m)j_e^2/4 B_0^2$.

\begin{figure}

\begin{center}
\includegraphics[width=1.0\columnwidth]{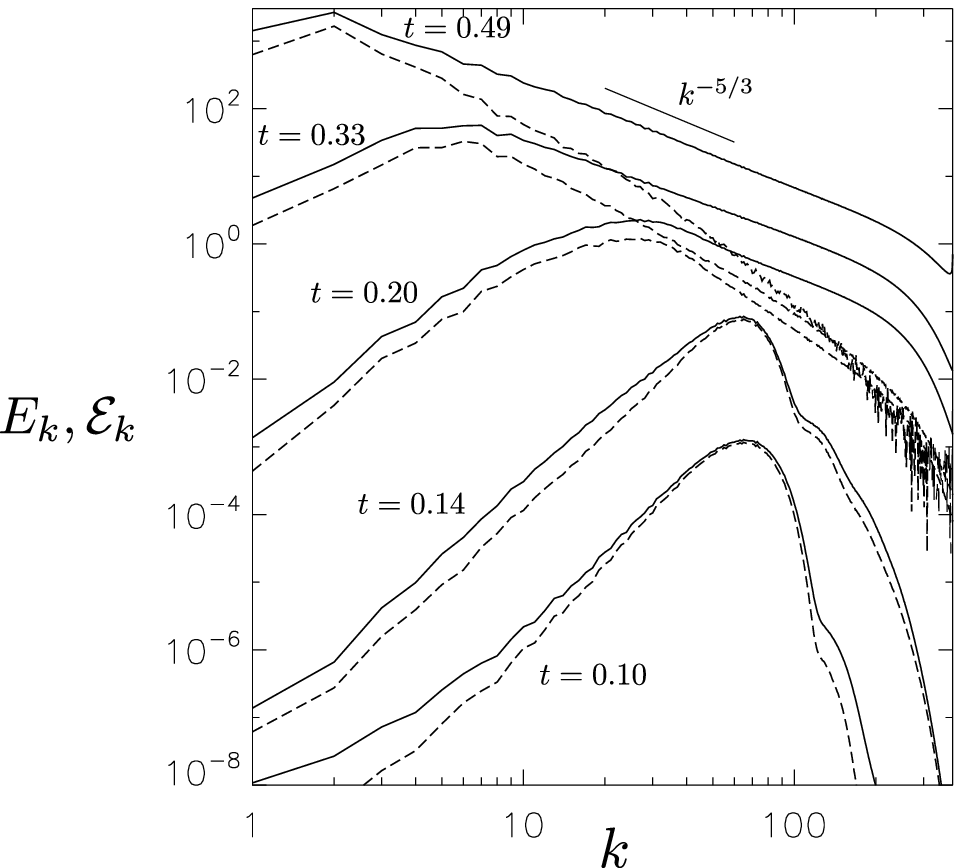}
\end{center}
\caption{Energy spectra (solid) and EMF spectra (dashed) at several moments of simulation time. We only take the vector component of EMF which is parallel to $j_e$, therefore the dashed spectrum is also the energy injection spectrum.
First two spectra feature linear mode growth with EMF spectrum proportional to the energy spectrum. Later, nonlinear stages show EMF driving dominant at the largest, energy
containing scale of the spectrum. Below energy-containing scale the energy injection becomes negligible and the spectrum exhibiting an ``inertial range'' $k^{-5/3}$ scaling. The approximate Kolmogorov scaling in the nonlinear regime
was reported earlier by \cite{rog2012}.}
\label{spectrum}
\end{figure}

The nonlinear stage takes over when nonlinear terms of Eqs. (\ref{dv}-\ref{ind}) become comparable with linear terms. This will happen when energy density exceeds characteristic
initial energy density $B_0^2/2$. Note that in the limit of ideal incompressible MHD statistically homogeneous system described by Eqs. (\ref{dv})-(\ref{ind}) has a single characteristic
energy density scale, $B_0^2$, timescale $1/j_e$, which corresponds to the linear growth rate, and lengthscale $B_0/j_e$, which corresponds to the wavelength of the most unstable mode.
If we conjecture that $B_0$ will become unimportant later in the nonlinear evolution, the system will only have a characteristic timescale $1/j_e$. It also turns out that dissipative effects can
be neglected as long as $B_0/j_e$ is above all dissipative scales.

Suppose at some moment of time during nonlinear evolution the spectrum of perturbations has an integral (outer) scale $L$. We will argue that perturbations on this scale
will be able to freely expand due to the Lorentz force, just like an unstable helical mode in the linear regime. We will also conjecture, by analogy with linear stage, that the product
${\bf j_e\cdot (v\times B)}$, i.e. the work of external current onto the fluid, will be proportional to energy. The difference with the linear stage is that the EMF spectrum will no longer
be proportional to the energy spectrum, rather it will peak at the integral scale. The result will be the growth of energy at the integral scale, the growth of the integral scale itself and the direct
energy cascade on scales below integral scale. From here on we introduce two dimentionless constants  that describe this process: the ratio of EMF to total energy (which conventionally
have the same Heaviside-Alfven units) $C_E$ and the fraction
of energy that goes into the direct cascade $C_D$. We expect $C_E$ and $C_D$ to be below unity. The whole spectrum of perturbations
at any moment of time will be determined only by the total energy $E$ and the integral scale $L$. The time evolution for these will be determined by

\begin{equation}
\frac{dE}{dt}=(1-C_D) C_E j_e E.
\end{equation}

\begin{figure}
\begin{center}
\includegraphics[width=0.8\columnwidth]{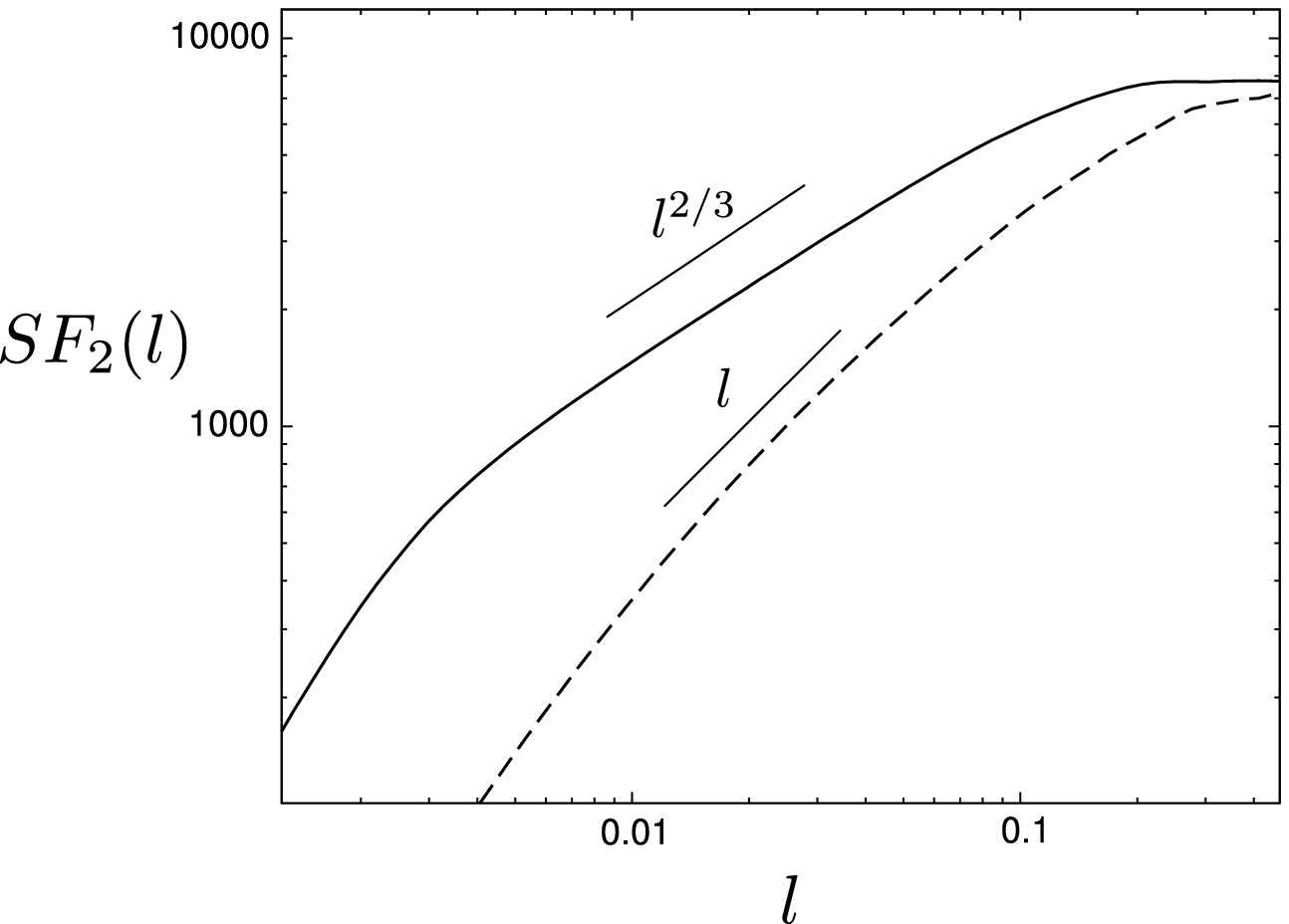}
\end{center}
\caption{Anisotropy of Bell-MHD turbulence below the energy containing scale at $t=0.49$. In this plot we show second order structure functions of density parallel (dashed) and perpendicular (solid) to the local mean magnetic field. The scalings
of $l^1$ and $l^{2/3}$ are expected from inertial range of strong MHD turbulence \cite{GS95}. Our measurement is grossly consistent with the Goldreich-Sridhar anisotropy.}
\label{sf}
\end{figure}

At the same time the zeroth law of turbulence, which states that at the sufficiently high Reynolds
numbers\footnote{This condition, in our case, will be satisfied as long as the dissipation scales are much smaller than all relevant scales of the problem,
including $1/k_d=B_0/j_e$. Given the expressions below, it is easy to show that as long as $\nu,\nu_m<B_0^2/j_e$, i.e. there is a positive growth rate
in the linear regime, the dissipation scale will be smaller than $1/k_d (B_0/B)^{1/2}$, i.e. this condition is always satisfied as long as there is a growth
of instability. Naturally, the small but finite dissipation effects are required for the turbulent cascade to terminate and produce heat.}
the dissipation will only depend on large-scale quantities, will read:

\begin{equation}
\frac{E^{3/2}}{L}=C_K C_D C_E j_e E. \label{zeroth}
\end{equation}

Here we have introduced Kolmogorov constant $C_K$. This will eventually give us the relation between the integral scale and the energy density:

\begin{equation}
L=E^{3/2}/(C_K C_D C_E j_e E))= E^{1/2}/(C_K C_D C_E j_e). \label{outer}
\end{equation}


\begin{figure}
\begin{center}
\includegraphics[width=0.9\columnwidth]{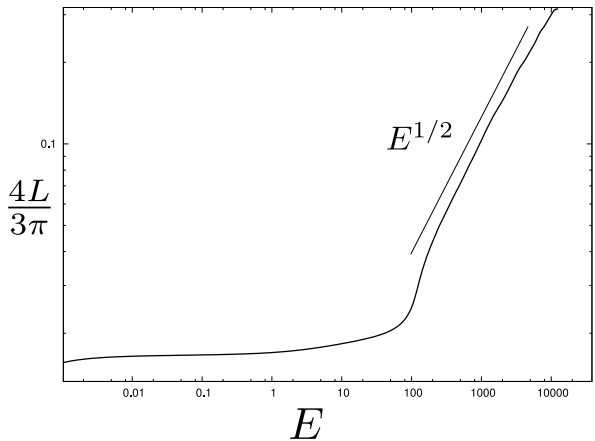}
\end{center}
\caption{The dependence of energy containing scale $L$ on energy density $E$. The linear growth regime features constant $L$, corresponding to the wavelength of the fastest growing mode. The nonlinear regime is characterized by the $l\sim E^{1/2}$
law, with both $E$ and $l$ growing exponentially in time. Outer scale $L$ is defined through spectrum as $(3\pi/4)\int k^{-1} E(k) dk/E$ \citep{Gotoh2002}.}
\label{l_vs_e}
\end{figure}

Since we expect equipartition between magnetic and kinetic energies, Eq.~(\ref{outer}) has a simple physical meaning. Since $E^{1/2} \sim B$, the fluctuations of the magnetic field
are such that the fluctuations of the total current density (plasma plus external) are of the order of the external current density itself, on the outer scale of turbulence. In other words, collisions
of expanding spirals from Bell instability almost fully randomizes total current, but its RMS value always stay around $j_e$.

The scalings above could be verified in numerics, in particular we observe that the energy density clearly continues to grow exponentially in the nonlinear stage, Fig.~\ref{energy}. 
And the outer scale $L$ indeed follows the above scaling (see Fig.~\ref{l_vs_e}). We can also directly verify that EMF drives turbulence mostly on the integral scale.

One of the possibilities brought up recently by \cite{rog2012} is that since the original Bell's instability generates helical states, the kinetic helicity of the resulting
turbulence will amplify any large-scale field on scale $l$, roughly at a rate $\alpha/l$, where $\alpha$ is due to the $\alpha$-effect of helical turbulence.
Our model does not explicitly deal with helicities, however it assumes that the mechanism of growth is essentially the same as in the linear regime -- i.e.
the expanding magnetic helixes. The difference is that this time the sign of helicity is determined by mutual orientation of $j_e$ and the local field, which becomes
increasingly isotropic as $B_0$ becomes energetically unimportant. In other words, we expect global fractional kinetic helicity to go to zero and the largest fluctuations
of helicity only present on the outer scale of turbulence $L$. Our conclusion, therefore, does not directly challenge the statement of \cite{rog2012} but amends it by saying that 
the growth of the field on scales which are much larger than $L$ is supressed due to kinetic helicity being averaged-out on such scales.

\section{Numerics}

We performed a series of numerical experiments to verify and refine the hypothethese stated above. We used incompressible pseudospectral
code which solves Eqs. (1-2) with extra dissipation terms which are used to regularize numerical solution. For dissipation terms
we used hyperdiffusivity of fourth order, equal in both $v$ and $B$, purely out of convenience: 1) the hyperdiffusivities allowed us to push $k_d$
closer to the Nyquist frequency $k_N$ without affecting linear growth rate too much, typically we used $k_d\approx 0.1 k_N$. This allowed us to have
greater scale separation between the box size and $1/k_d$ to study self-similar nonlinear behavior for a greater range of scales and energies; 2) the dissipation
rate varied by several orders of magnitude during the simulation and having second-order diffusivities would have required constantly tuning diffusion coefficients
to keep the simulation well-resolved. Regarding further details of the code architecture, accuracy and implementation we refer to our previous publications,
e.g. \citet{BL09a}. It is also worth noting that the code preserved scaling symmetries of the incompressible MHD equations, e.g. rescaling time in proportion with $1/j_e$,
while keeping $j_e/B_0$ constant would result in exactly the same evolution.

We performed a series of six $384^3$ experiments changing the value
of initial magnetic field $B_0$ between values of 0.5, 1.0 and 2.0 and changing current between 40 and 20. Due to the symmetries of the dynamical equations,
we expected simulations with the same $j_e/B_0$ ratio to exhibit the same behavior, assuming that the timescale was properly rescaled to $1/j_e$. This was indeed the case.
Also, as long as the scale $1/k_d$ was well separated from both dissipation scale and the cube size scale, the behaviour was similar for all simulations,
assuming the same rescaling. Due to the fact that incompressible MHD are scale-free, the only true physical parameter of the problem
is $j_e$ and since it has units of $1/s$, it simply designates the only available timescale for the problem. In other words, the evolution is expected to be universal
as long as the dissipation coefficients are small and the box size is large enough.
Keeping this in mind we performed a single $1152^3$ large-scale simulation to verify the scaling in Eq.~(\ref{outer}). In this simulation we chose $j_e=120$, $B_0=1$,
so that $k_d=60$ and is well-separated from dissipation scale at around $k=300$. Figures 1-5 present measurements from this simulation.
The total energy evolution is presented on Fig.~\ref{energy}, while spectra of total energy and the EMF along ${\bf j_e}$ are presented
on Fig.~\ref{spectrum}. A few comments on these results are in order. First of all, the nonlinear regime exhibits a pure exponential growth, over at least a couple
of order of magnitude in energy, which is fully consistent with our model and implicitly verifies that the external current density $j_e$ determines the only relevant timescale of the problem.
The coefficient of reduction of the growth rate, $0.132$, that we conjectured to be equal to the $(1-C_D) C_E$ product indeed seems to be universal among several
simulations with different $j_e$, $B_0$ and numerical resolutions. We observe the transition to nonlinear stage at approximately the same level of $B_{\rm rms}/B_0\sim 10$
as previous simulations, e.g., \cite{rog2012}.

\begin{figure}
\begin{center}
\includegraphics[width=0.9\columnwidth]{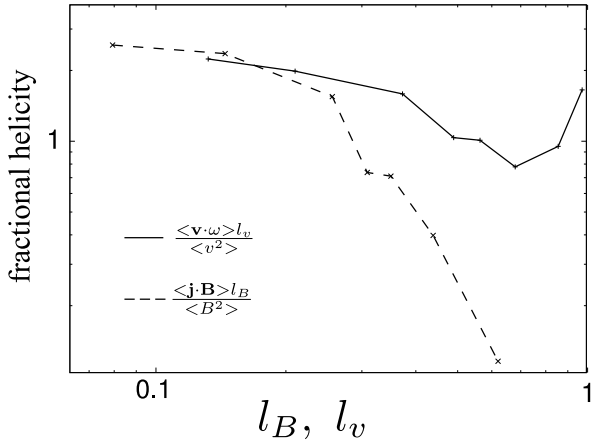}
\end{center}
\caption{Fractional kinetic helicity vs kinetic scale (solid) and fractional current helicity vs magnetic scale (dashed).
While fractional kinetic and current helicities are large in the linear mode growth, they decrease during nonlinear growth,
which is due to randomization of magnetic field direction with respect to the external current direction. When the other scale $L$ approaches the box size,
however, the kinetic helicity starts growing again, while current helicity changes sign.}
\label{hel}
\end{figure}

Furthermore, the spectra of energy and EMF, Fig.~\ref{spectrum}, seem to support our conjectures about the scales at which the Bell-MHD turbulence is driven.
For the linear growth regime, represented on Fig.~\ref{spectrum} by spectra at t=0.1 and t=0.14, the driving EMF spectrum is basically proportinal to the energy spectrum,
while for nonlinear  growth regime at t=0.2, 0.33 and 0.49 the EMF is peaked on the outer scale, supporting our conjecture about large-scale driving. In fact, in order to be
important on smaller scales $l$ and to interfere with the energy cascade through scales the EMF has to be at least scale-independent or to grow with smaller scales,
which is clearly not the case. From the reduced ratio of EMF to energy in the nonlinear regime, we derive $C_E\approx 0.58$ and using  $(1-C_D) C_E=0.132$ we have
the fraction of energy captured by the direct cascade and dissipated as $C_D=0.77$.

It is also interesting to study the statistical properties of the nonlinear Bell-MHD turbulence and see if they are similar to the ordinary direct-cascade MHD turbulence
driven on large scales. We measured second-order structure function $SF(l)$ parallel and perpendicular to the local magnetic field, which are supposed to scale as $l$
and $l^{2/3}$ correspondingly, according to the standard model of strong MHD turbulence by \cite{GS95}. We found that the structure function measurement grossly
consistent with the theory, see Fig.~\ref{sf}. One of the important consequences of this is that due to local anisotropy, the Bell-generated turbulence will be inefficient
at scattering particles with gyroradii much smaller than the outer scale $L$, see, e.g., \cite{chandran2000,yan2002}, just like the driven MHD turbulence.

As the instability enters the nonlinear stage, the outer scale quickly grows, in fact, exponentially in time. Fig.~\ref{l_vs_e} investigates the dependence
of the outer scale on the mean energy density and finds that indeed outer scale grows as the square root of energy density as Eq.~(\ref{outer}) suggests.
We found that $L\approx 0.97 E^{1/2}/j_e$. Assuming equpartition between magnetic and kinetic energies which is indeed approximately satisfied,
and going back to CGS units in the above relation, we obtain $B_{\rm rms}{\rm (CGS)} \approx 4\pi j_e{\rm (CGS)} L{\rm (CGS)}/c$.  

Let us discuss the role of kinetic and current helicities. Starting with the linear stage of the instability, when the plasma current equals
approximately ${\bf -j_e}$, the system has large current helicity $-\bf j_e\cdot B_0$ and the linearly growing mode also has large kinetic
helicity of the same sign. As the instability enters the nonlinear stage, however, one would expect that the infuence of the mean field
$B_0$ will decrease. The helical term $-\bf j_e\cdot B_0$ will be relatively less important at later stages of the nonlinear evolution,
while the fluctuating $-\bf j_e\cdot B(r)$ will be more important. At each snapshot of the nonlinear evolution, given the outer
scale of Bell turbulence $L(t)$ at this time, the most contribution to both the energy growth and kinetic and current helicities will come from
scale $L(t)$ because the scales below $L(t)$ represent direct cascade. However, averaged over the system size, which is expected to be much larger
than $L(t)$, the total helicity should be relatively small, as at the each eddy of size $L(t)$ the helical terms are determined by randomly oriented
$-\bf j_e\cdot B(r)$. The high local kinetic helicity on scale $L(t)$, therefore, will result in a growth of larger scale field on the next available largest scale,
figuratively speaking $2L(t)$, but will not result in a growth on much larger scales, because at those scales the $-\bf j_e\cdot B(r)$ is randomized.

In order to check this hypothesis we calculated global kinetic and current helicities as a function of time. Fig.~\ref{hel} shows dimentionless fractional
helicities as a function of magnetic and kinetic scale. While in the linear stage $-\bf j_e\cdot B_0$ term is dominant and both fractional
current and kinetic helicities are large, as the instability progresses to larger scales, global fractional helicities decrease. This is because
the large helicities produced locally on scale $L$ are of random sign and are averaged out on the scale of the box. It is interesting, however,
that when the outer scale $L$ approached the box size the fractional kinetic helicity started growing again. We believe that when the system
reaches the box size it chooses the dominant global helical state and sticks to it.
As we believe that the supernova precursor size is always much larger than
the outer scale $L$, this should be interpreted as an artificial numerical effect, associated with the finite box size.
 
Lastly, to further challenge the idea of prevalent infulence of initial current helicity, we investigated the case with zero mean field $B_0$.
This situation is special in a sense that the system
did not have any preferred helicity to begin with. The instability, therefore has to grow starting with low-level magnetic noise with which we seeded the simulation.
In this case the fast linear growth was not observed, however, after some period of slower growth the simulation entered nonlinear stage.
In this nonlinear stage the growth rate was consistent with $0.132 j_e$, the same rate that we observed in the nonlinear stage of all helical cases. 
  
\section{Discussion}
Our estimate of outer scale Eq.~(\ref{outer}) can be readily converted to the estimates of the maximum energy of accelerated particles, assuming that current is due
to all CRs drifting with the average speed of the shock $v_s$:

\begin{equation}
E_{\rm max}= \frac{v_s}{c} eB_{\rm rms}L \approx \frac{B_{\rm rms}^2}{4\pi n_{\rm CR}}
\end{equation}

That is, the {\it maximum} particle energy equals to the average magnetic energy per particle.
The estimate of maximum magnetic energy density in terms of cosmic ray energy density $ U_{\rm CR}$ was gien by \citet{bell2004} as $(v_s/c) U_{\rm CR}$.
Using this estimate we get $E_{\rm max}=2 (v_s/c) U_{\rm CR}/n_{\rm CR}$, i.e. $E_{\rm max}$ is much smaller than the average cosmic ray energy.
The estimate of \citet{voelk2005} gives $B_{\rm rms}^2/(8\pi)=1.7\times 10^{-2} \rho v_s^2$, based on observational data. Using this estimate and
assuming that $U_{\rm CR}=10^{-1} \rho v_s^2$, $E_{\rm max}=0.34 U_{\rm CR}/n_{\rm CR}$. The average energy of CRs $U_{\rm CR}/n_{\rm CR}$
is expected to be low, around 1 GeV, however in any case it should be lower than maximum energy, not vice versa. Furthermore, if we assume that some
of the current is due to cosmic rays escaping with a speed of light, i.e. $j_e>e v_s n_{\rm CR}$ our estimate of $E_{\rm max}$ will be even lower, somewhat
counter-intuitively, because our Eq.~(\ref{outer}) suggests that the saturation scale of $L$ will be inversely proportional to $j_e$. 
All in all, assuming that classic Bell's mechanism increases maximum energy of accelerated particles, we came to a contradiction\footnote{In the Equation (6)
we used the condition of efficient scattering for high energy particles, so that the Larmor radius $r_L$ should be smaller than $v_s L/c$. We do not expect
the cosmic rays precursor to be much thicker that $L$ in the pure Bell instability case, because the precursor thickness will be determined by the saturation of
the instability itself and the efficient scattering of low-energy particle that should happen on scales smaller than $L$. However, if one believes that the Bell-CR precursor
is much thicker, one may use the other formula for the scattering rate for the case $r_L\gg L$. This should give the maximum larmor radius of accelerated particles
as $r_{\rm max}=(v_s L_p L/c)^{1/2}$, where $L_p$ is the precursor thickness. However, assuming maximum field strength of 100$\mu$G, even fairly
large $L_p\approx 0.1$pc still gives maximum energies of around 2 TeV, way below the knee of the CR spectrum.}.
It seems that without considering the feedback of magnetic perturbations on cosmic rays, using just classic current instability will not
result in decreased diffusion of cosmic rays on the high-energy end. When considering feedback, however, treating cosmic rays as homogeneous current
is not sufficient and gyroresonance effects should be included.

Let us compare the $L\sim E^{1/2}$ dependence produced by Bell's mechanism with other magnetic field generation mechanisms.
The small-scale dynamo have a $L\sim E^{3/2}$ dependence, with $E$ being the energy density of the magnetic field \citep{B12a}, thus for these two
mechanisms the product of $LB$, which determines the scattering of highest energy particles, will be $E^1$ and $E^2$ correspondingly. Given that
the magnetic energy density $E$ is bound on energetic grounds, and also constrained by observations \citep{vink2003,cassam2007, voelk2005},
the $LB \sim E^2$ seems more favorable than $LB \sim E$.

Prior numerical work investigating linear and nonlinear stages of Bell's instability includes \citet{bell2004,zirakashvili2008,riquelme2009,rog2012}.
Transition to the nonlinear regime has happened at about the same level of perturbations in \citet{bell2004,zirakashvili2008,rog2012} and this investigation.
All fluid simulations have shown some growth at the nonlinear stage, but the exponential nature of this growth
was less evident compared to our Fig~1. The above simulations were interpreted in various ways using phenomenological argumentation.
For example, \citet{bell2004} suggested that the magnetic spectrum will be defined by scale-wise equilibrium of magnetic tension, which will result in
a power spectrum of $k^{-3}$. \citet{rog2012} noticed that the spectrum is shallower than that, around $-5/3$, due to the presence of the direct
cascade. While all of the above papers mention the growth of the outer scale in the nonlinear regime, only \citet{rog2012} made quantitative predictions
regarding such growth. This paper was the first to measure anisotropy in Bell-MHD turbulence. 
Plasma simulations of \citet{riquelme2009} qualitatively confirmed current instability and identified some
differences compared to the fluid case. The detailed comparison with this paper would be outside the scope of this presentation.   

The key properties of the Bell-MHD turbulence we have uncovered, such as Eq.~(\ref{outer}) suggest that, when considering including the feedback
on cosmic rays and/or making conjectures about their possible roles in cosmic ray acceleration, one has to be very careful in adopting the
spectrum and amplitude of such turbulence. This is because this turbulence will strongly affect the properties of lower-energy CRs, i.e. the same
particles that mostly contribute to current. Another source of uncertainty would be undestanding possible effects of compressible turbulence.
Our picture of growing magnetic helixes on the outer scale $L$ due to interaction with local external current does not expicitly depend on fluid pressure,
however, and the zeroth law of turbulence Eq.~(\ref{zeroth}), to the best of our knowledge, is supposed to be correct even in supersonic turbulence.
Furthermore, since our results indicate that the $C_D$ fraction of total energy goes into the direct cascade and dissipates into heat, the sonic Mach number of Bell-MHD
turbulence can be estimated as $M_s\approx \sqrt{(1-C_D)/2C_D} \approx 0.4$. Although qualitatively the picture of nonlinear stage is not supposed to change,
we might expect the coefficients $C_D$, $C_E$ and $C_K$ to somewhat vary in the compressible case.

Our results are applicable not only to supernova remnants problem, but to any conductive fluid with external current, e.g., plasma with rigid wires
embedded in it and the current driven by an external voltage. The quick exponential growth of both energy and fluid EMF suggests that external
energy source will only be able to maintain such current for a limited amount of time. 

Several pathways are available to study back-reaction to cosmic rays. One of the approaches is the bottom-up plasma simulations, e.g., by \citet{riquelme2009}
who tried to capture scaling behaviours that may be extended to larger scales. The full treatment of nonlinear streaming instability \citep{diamond2007} is the most general
and desired approach, but has significant analytical difficulties. The path to explore the long wavelength dynamics while keeping Bell's instability 
intact \citep{bykov2011,rog2012} has also been popular recently. The entirely different approach to get turbulence driven on large scales by preexisting
density inhomogeneities and the non-barotropic cosmic ray pressure was suggested in \citet{BJL09} and currently being pursued by \citet{Drury2012, Bruggen2013}
and the authors of this paper. We believe that large-scale three-dimensional  MHD-particle simulations could shed light on this extremely complex theoretical problem.

\section{Acknowledgments}
AB is grateful to Mischa Malkov, Axel Brandenburg, Tony Bell, Vladimir Zirakashvili, Pat Diamond and Alex Schekochihin for useful discussions. 
The work is supported by the LANL/LDRD program and the DoE/Office of Fusion Energy Sciences through CMSO. 
Computing resources at LANL are provided through the Institutional Computing Program.
 
\

\def\apj{{\rm ApJ}}           
\def\apjl{{\rm ApJ }}          
\def\apjs{{\rm ApJ }}          
\def\grl{{\rm GRL }}
\def\aap{{\rm A\&A } }
\def\mnras{{\rm MNRAS } }
\def\physrep{{\rm Phys. Rep. } }               
\def\prl{{\rm Phys. Rev. Lett.}} 
\def\pre{{\rm Phys. Rev. E}} 
\bibliography{all}

\end{document}